\documentclass[a4paper,12pt]{article}
\usepackage[utf8]{inputenc}
\usepackage{amsmath, amssymb, comment, hyperref, cite, graphicx, authblk}
\usepackage{float}

\usepackage{anysize}
\marginsize{2.0cm}{2.0cm}{2.0cm}{2.0cm}
\begin{document}

\title{Green's functions for higher order nonlinear equations \\ {\normalsize Case studies for KdV and Boussinesq equations}}

\author[1]{Marco Frasca}

\author[2, 3]{Asatur Zh. Khurshudyan\footnote{Email: khurshudyan@mechins.sci.am}}

\affil[1]{\small Via Erasmo Gattamelata 3, 00176 Rome, Italy}
\affil[2]{\small Department on Dynamics of Deformable Systems and Coupled Fields, Institute of Mechanics, National Academy of Sciences of Armenia, Yerevan, Armenia}
\affil[3]{\small Institute of Natural Sciences, Shanghai Jiao Tong University, Shanghai, China}

\date{}

\maketitle

\begin{abstract}

The well-known Green's function method has been recently generalized to nonlinear second order differential equations. In this paper we study possibilities of exact Green's function solutions of nonlinear differential equations of higher order. We show that, if the nonlinear term satisfies a generalized homogeneity property, then the nonlinear Green's function can be represented in terms of the homogeneous solution. Specific examples and a numerical error analysis support the advantage of the method. We show how, for the Bousinesq and Kortweg-de Vries equations, we are forced to introduce higher order Green functions to obtain the solution to the inhomogeneous equation. The method proves to work also in this case supporting our generalization that yields a closed form solution to a large class of nonlinear differential equations, providing also a formula easily amenable to numerical evaluation.


\end{abstract}

\section{Introduction}

From precise applied modelling viewpoint, nonlinear models are more desirable than the linear ones. Nevertheless, the analysis of nonlinear models is far too complicated compared with that of linear models. For them, the rigorous analysis is extremely complicated (sometimes even impossible) and the numerical methods require much computational efforts. However, for some specific systems it is still possible to employ (semi-)analytical methods. One of the powerful analytical methods is the nonlinear Green's function method developed by Frasca about a decade ago for specific applications in quantum field theory. It has been established by Frasca that the solution of the second order quasi-linear ODE
\begin{equation}\label{gennonlin}
\frac{d^2 w}{d t^2} + N\left(w\right) = f\left(t\right), ~~ t > 0,
\end{equation}
with a generic non-linearity $N$ and a given source function $f$, under appropriate Cauchy conditions admits the following short time expansion of the general solution \cite{Frasca:2005sx, Frasca2007, Frasca2008}:
\begin{equation}\label{shorttime}
w\left(t\right) = a_0 \int_0^t G\left(t - \tau\right) f\left(\tau\right) {\rm d}\tau + \sum_{k = 1}^\infty a_k \int_0^t \left(t - \tau\right)^k G\left(t - \tau\right) f\left(\tau\right) {\rm d}\tau,
\end{equation}
where unknowns $a_k$, $k = 0, 1, 2, \dots,$ are determined in terms of the quantities $w^{\left(k\right)}\left(0\right)$.

Here $G$ is the solution of the differential equation
\begin{equation}\label{Greeneq}
\frac{d^2 G}{d t^2} + N\left(G\right) = s \delta\left(t\right),~~ t > 0,
\end{equation}
under corresponding Cauchy conditions, where $\delta$ is the Dirac distribution. Due to the similarity with the linear case, $G$ is referred to as nonlinear Green's function of (\ref{gennonlin}).

Furthermore, it has been shown in \cite{Frasca2018(1)} that the short time expansion (\ref{shorttime}) also holds true in the case when the nonlinear term depends on the first derivative of the unknown function:
\[
N = N\left(\frac{d w}{d t}, w\right).
\]

It is noteworthy that even the first order term in the short time expansion above, i.e.,
\begin{equation}\label{Frascaapprox}
w\left(t\right) \approx a_0 \int_0^t G\left(t - \tau\right) f\left(\tau\right) {\rm d}\tau,
\end{equation}
provides a numerical approximation of differential equations consistent with the numerical solution obtained by the well-known method of lines, while higher order terms contribute to the error corrections. Note also that in previous papers \cite{Frasca:2005sx, Frasca2007, Frasca2008} $a_0 = 1$, while in the papers \cite{Khurshudyan2018, Khurshudyan2018AMP, Khurshudyan2018IJMPC}, $a_0$ is introduced to minimize the approximation error.

In this paper we address the case of higher order equations. Assuming that the nonlinear equation is linear in the highest order derivative, a representation formula for the nonlinear Green's function in terms of the corresponding homogeneous solution will be given. In order for this to work, the nonlinear term must satisfy a certain relation very similar to the homogeneity property of functions. Some particular non-linearities satisfying this property have been obtained in \cite{Frasca2018}. The most challenging part in this study is that the majority of higher order equations, unlike second order nonlinear equations, are not solved exactly (explicitly or implicitly). Specific examples of higher order nonlinear PDEs are considered to demonstrate the Green's function determination procedure, including Kortweg-de Vries and Boussinesq equations. In this paper we will also consider higher-order Green functions that are needed when one can solve the equation for Dirac distribution derivatives. These will prove relevant both for the Kortweg-de Vries and the  Bousinesq equations. Numerical error analysis supports the advantages of the method we discuss in giving an explicit successful solution to such a class of equations and for numerical evaluation purposes.

\section{Representation of the nonlinear Green's function}

\subsection{Nonlinear Green's function for second order equations}

In \cite{Frasca2018, Frasca2018(1)}, it has been proved that under proper restrictions on the nonlinear term, it is possible to derive a simple representation for the nonlinear Green's function. More specifically, consider the following second order ODE:
\begin{equation}\label{nonlinODE}
\frac{d^2 w}{d t^2} + N\left(\frac{d w}{d t}, w\right) = f\left(t\right), ~~ t > 0.
\end{equation}
Then, as soon as
\begin{equation}\label{Ncondderiv}
    N\left(\theta \cdot \frac{d w}{d t}, \theta \cdot w\right) = \theta\left(t\right) \cdot N\left(\frac{d w}{d t}, w\right),
\end{equation}
the Green's function of (\ref{nonlinODE}) admits the following representation:
\begin{equation}\label{Greenrep}
    G\left(t\right) = \theta\left(t\right) w_0\left(t\right),
\end{equation}
where $w_0$ is the general solution of the corresponding homogeneous equation
\[
\frac{d^2 w}{d t^2} + N\left(\frac{d w}{d t}, w\right) = 0,
\]
subject to the following Cauchy conditions:
\begin{equation}\label{nonhomCauchy}
w\left(0\right) = 0, ~~ \frac{d w}{d t}\bigg|_{t = 0} = s.
\end{equation}
Here, $\theta$ is the Heaviside function. For particular non-linearities satisfying (\ref{Ncondderiv}) see \cite{Frasca2018}.

Higher-order Green functions can also be considered. This case has been treated in \cite{Frasca2007(1)}. We assume that one has the solutions to equations like
\begin{eqnarray}
\frac{d^2 G_1}{d t^2} + N\left(\frac{d G_1}{d t}, w\right) &=& s_1\delta'(t) \nonumber \\
\frac{d^2 G_2}{d t^2} + N\left(\frac{d G_2}{d t}, w\right) &=& s_2\delta''(t) \nonumber \\
&\vdots&. \nonumber
\end{eqnarray}
and we miss the knowledge of the solution of  eq.~(\ref{Greeneq}). The leading order solution to eq.~(\ref{nonlinODE}) can be immediately written down also for these cases as
\begin{eqnarray}
w\left(t\right) &\approx& a_0^{(1)} \int_0^t G_1\left(t - \tau\right) {\rm d}\tau\int_0^{\tau}f\left(\tau'\right) {\rm d}\tau' \nonumber \\
w\left(t\right) &\approx& a_0^{(2)} \int_0^t G_2\left(t - \tau\right) {\rm d}\tau\int_0^{\tau} {\rm d}\tau'\int_0^{\tau'}f\left(\tau''\right) {\rm d}\tau'' \nonumber \\
&\vdots&. \nonumber
\end{eqnarray}
We will apply this to the Kortweg-de Vries and Bousinesq equations.

\subsection{Nonlinear Green's function for higher order equations}

It can be shown that, under proper assumptions, the representation (\ref{Greenrep}) holds also for higher order equations. Consider the following ODE:
\begin{equation}\label{HOnonlinODE}
\frac{d^n w}{d t^n} + N\left(\frac{d^{n - 1} w}{d t^{n - 1}}, \dots, \frac{d w}{d t}, w\right) = f\left(t\right), ~~ t > 0,
\end{equation}
for arbitrary $2 < n \in \mathbb{N}$. Let $w_0$ be the general solution of the following Cauchy problem:
\begin{equation}\label{HOnonlinODEhomogen}
\frac{d^n w}{d t^n} + N\left(\frac{d^{n - 1} w}{d t^{n - 1}}, \dots, \frac{d w}{d t}, w\right) = 0, ~~ t > 0,
\end{equation}
\begin{equation}\label{HOCauchy}
w\left(0\right) = \frac{d w}{d t}\bigg|_{t = 0} = \dots = \frac{d^{n - 2} w}{d t^{n - 2}}\bigg|_{t = 0} = 0, ~~ \frac{d^{n - 1} w}{d t^{n - 1}}\bigg|_{t = 0} = s.
\end{equation}
Then, taking into account that in the sense of distributions,
\[
\frac{d \theta}{d t} = \delta\left(t\right),
\]
\[
\frac{d^n \left(\theta \cdot w\right)}{d t^n} = \sum_{k = 1}^{n} \frac{d^{k - 1} w}{d t^{k - 1}}\bigg|_{t = 0} \delta^{\left(n - k\right)}\left(t\right) + \theta\left(t\right) \frac{d^n w}{d t^n},
\]
we see that as soon as $N$ is a generalized homogeneous function in the following sense:
\begin{equation}\label{HONmultipl}
N\left(\theta \cdot \frac{d^{n - 1} w}{d t^{n - 1}}, \dots, \theta \cdot \frac{d w}{d t}, \theta \cdot w\right) = \theta\left(t\right) \cdot N\left(\frac{d^{n - 1} w}{d t^{n - 1}}, \dots, \frac{d w}{d t}, w\right),
\end{equation}
then the function
\begin{equation}\label{HOGreenrep}
    G\left(t\right) = \theta\left(t\right) w_0\left(t\right),
\end{equation}
satisfies the following equation:
\begin{equation}\label{HOGreenODE}
\frac{d^n G}{d t^n} + N\left(\frac{d^{n - 1} G}{d t^{n - 1}}, \dots, \frac{d G}{d t}, G\right) = s \delta\left(t\right).
\end{equation}
Thus, (\ref{HOGreenrep}) is the nonlinear Green's function of (\ref{HOnonlinODE}).

It is noteworthy that many important hierarchies of non-linearities satisfying (\ref{HONmultipl}) can be constructed using the solutions obtained in \cite{Frasca2018}.

\section{Nonlinear Green's function for some higher order equations}

\subsection{Kor\-te\-weg-de Vries equation}

An interesting case, due to its widespread applications in the study of solitons, is that of the Korteweg-de Vries equation
\[
\frac{\partial \tilde{w}}{\partial t} + \frac{\partial^3 \tilde{w}}{\partial x^3} - 6 \tilde{w} \frac{\partial \tilde{w}}{\partial x} = 0.
\]

The traveling wave ansatz
\[
\tilde{w}\left(x, t\right) = w\left(x - c t - a\right) := w\left(\zeta\right),
\]
where $c, a \in \mathbb{R}$, reduces the above PDE to the following third order ODE:
\[
-c \frac{d w}{d \zeta} + \frac{d^3 w}{d \zeta^3} - 6 w \frac{d w}{d \zeta} = 0.
\]
Apparently, this equation can be reduced to a second order equation by taking into account that it is equivalent to the following equation:
\[
\frac{d}{d \zeta} \left[ \frac{d^2 w}{d \zeta^2} - 3 w^2 - c w \right] = 0.
\]
Integration of the latter with respect to $\zeta$ once implies
\begin{equation}\label{KdVODEred}
\frac{d^2 w}{d \zeta^2} - 3 w^2 - c w = c_0,    
\end{equation}
where $c_0$ is an arbitrary constant. Assume that $c_0 = 0$. Then, it is easy to compute that the function
\begin{equation}\label{GreenKdV}
G_1\left(\zeta\right) = -\frac{c}{2} \theta\left(\zeta\right) \operatorname{sech}^2\left[\frac{\sqrt{c}}{2}\zeta\right]
\end{equation}
satisfies the following nonlinear ODE:
\[
\frac{d^2 G_1}{d \zeta^2} - 3 G_1^2 - c G_1 = -\frac{c}{2} \delta'\left(\zeta\right).
\]
In the terminology of \cite{Frasca2007(1)}, $G_1$ is the order one Green's function of the Kortweg-de Vries equation (\ref{KdVODEred}). This means, that its general solution has the leading order term as follows:
\[
w\left(\zeta\right) = a_0 \int_0^\zeta G_1\left(\zeta - \eta\right) F\left(\eta\right) {\rm d}\eta,
\]
with
\[
F\left(\zeta\right) = \int_0^{\zeta} f\left(z\right) {\rm d}z,
\]
being $f$ the source term.

\subsection{Quadratic non-linearity}

Consider the following fourth order PDE \cite{Polyanin2012}:
\begin{equation}\label{fourthorderPDE}
\frac{\partial^2 w}{\partial t^2} = \frac{\partial^4 w}{\partial x^4} + w \frac{\partial^2 w}{\partial x^2}.
\end{equation}
The traveling wave ansatz
\[
\tilde{w}\left(x, t\right) = w\left(x - v t\right) := w\left(\zeta\right),
\]
reduces the above PDE to the following fourth order ODE:
\begin{equation}\label{fourthorderODE}
\frac{d^4 w}{d \zeta^4} + \left(w\left(\zeta\right) - v^2\right) \frac{d^2 w}{d \zeta^2} = 0.
\end{equation}
Its Green's function admits the representation (\ref{Greenrep}). However, $w_0$ is not found exactly. Approximate form of $w_0$ can be obtained by, e.g., the Adomian decomposition method \cite{Adomian1994}, power series solution method, etc., can be used instead.

Then, the nonlinear Green's function provides the following representation for the travelling wave solution of forced version of (\ref{fourthorderPDE}):
\[
\tilde{w}\left(x, t\right) = w\left(x - v t\right) = \sum_{k = 0}^\infty a_k \int_0^{x - v t} \left(x - v t - \zeta\right)^k G\left(x - v t - \zeta\right) f\left(\zeta\right) {\rm d}\zeta,
\]
where $f$ is the forcing term.

\subsection{Boussinesq equation}

Another example of a higher order equation having widespread applications in physics and specifically in hydrodynamics, is the Boussinesq equation
\[
\frac{\partial^2 \tilde{w}}{\partial t^2} + \frac{\partial}{\partial x} \left[ \tilde{w} \frac{\partial \tilde{w}}{\partial x}\right] + \frac{\partial^4 \tilde{w}}{\partial x^4} = 0.
\]
The traveling wave {\sl ansatz}
\[
\tilde{w}\left(x, t\right) = w\left(x - v t\right) := w\left(\zeta\right),
\]
reduces the Boussinesq equation to the following fourth order ODE:
\[
\frac{d^4 w}{d \zeta^4} + v^2 \frac{d^2 w}{d \zeta^2} + \frac{1}{2} \frac{d^2 w^2}{d \zeta^2} = 0.
\]
A way of proceeding further may be the integration of this equation with respect to $\zeta$ twice. The resulting equation is a second order nonlinear non-homogeneous equation with a quadratic non-linearity and a source term linear in $\zeta$. Such an equation is studied in \cite{Khurshudyan2018AMP}.

On the other hand, the reduced equation admits cnoidal and snoidal solutions \cite{Chen2007}. In our case these can be written down as
\[
w\left(\zeta\right) = -\frac{3c^2 v^2}{1 + c^2} \operatorname{sn}^2\left(\frac{v \zeta}{2\sqrt{1 + c^2}} + \phi, c\right)
\]
and
\[
w\left(\zeta\right) = \frac{3c^2 v^2}{1 - 2c^2} \operatorname{cn}^2\left(\frac{v \zeta}{2\sqrt{1 - 2c^2}} + \phi, c\right),
\]
that has a singular point at $c^2 = 1/2$ where a change in behavior is expected. In these equations, $c$ and $\phi$ are arbitrary integration constants.

For our aims, the Green's function can be defined using the snoidal solution and we write down
\[
G_1\left(\zeta\right) = -\theta\left(\zeta\right) \frac{3c^2 v^2}{1 + c^2} \operatorname{sn}^2\left(\frac{v \zeta}{2\sqrt{1 + c^2}} + \phi, c\right),
\]
which is an order one Green's function as it solves the equation
\[
\frac{d^4 G_1}{d \zeta^4} + v^2 \frac{d^2 G_1}{d \zeta^2} + \frac{1}{2} \frac{d^2 G_1^2}{d \zeta^2} = -\frac{3}{4}\frac{c^2 v^4}{(1 + c^2)^2}\delta'\left(\zeta\right).
\]
This yields again a solution to the non-homogeneous Boussinesq equation provided we write it, at the leading order, as
\[
w\left(\zeta\right) = a_0 \int_0^\zeta  G_1\left(\zeta - z\right) F\left(z\right) {\rm d}z,
\]
with
\[
F\left(\zeta\right) = \int_0^{\zeta} f\left(z\right) {\rm d}z,
\]
being $f$ the source term. This implies that one can consider also $G_2$, $G_3$ and higher order Green's functions with higher derivatives of the Dirac distribution and obtain anyway the solution of the given equation. Therefore, integrability of the source term up to the corresponding order must be granted.

\section{Numerical error analysis}

In this section we numerically quantify the approximation error of some of the solutions obtained above. To this aim, we introduce the following logarithmic error function:
\[
\operatorname{Er}_1\left(\zeta; N\right) = \ln\left|w_{\rm Green's}^N\left(\zeta\right) - w_{\rm MoL}\left(\zeta\right)\right|,
\]
where $w_{\rm Green's}^N$ is the partial sum of the short time expansion (\ref{shorttime}) for a finite $N \in \mathbb{N}$; $w_{\rm MoL}$ is the numerical solution obtained by means of the well-known method of lines \cite{Schiesser1991}. We also aim to quantify the contribution of the higher order terms of the Frasca's short time expansion (\ref{shorttime}) by means of the logarithmic error
\[
\operatorname{Er}_2\left(\zeta; N_1, N_2\right) = \ln\left|w_{\rm Green's}^{N_1}\left(\zeta\right) - w_{\rm Green's}^{N_2}\left(\zeta\right)\right|,
\]
where $N_1 < N_2$ are the approximation orders.

\subsection{Kortweg-de Vries equation}

\begin{figure}[H]
\centerline{\includegraphics[width = 3.2in]{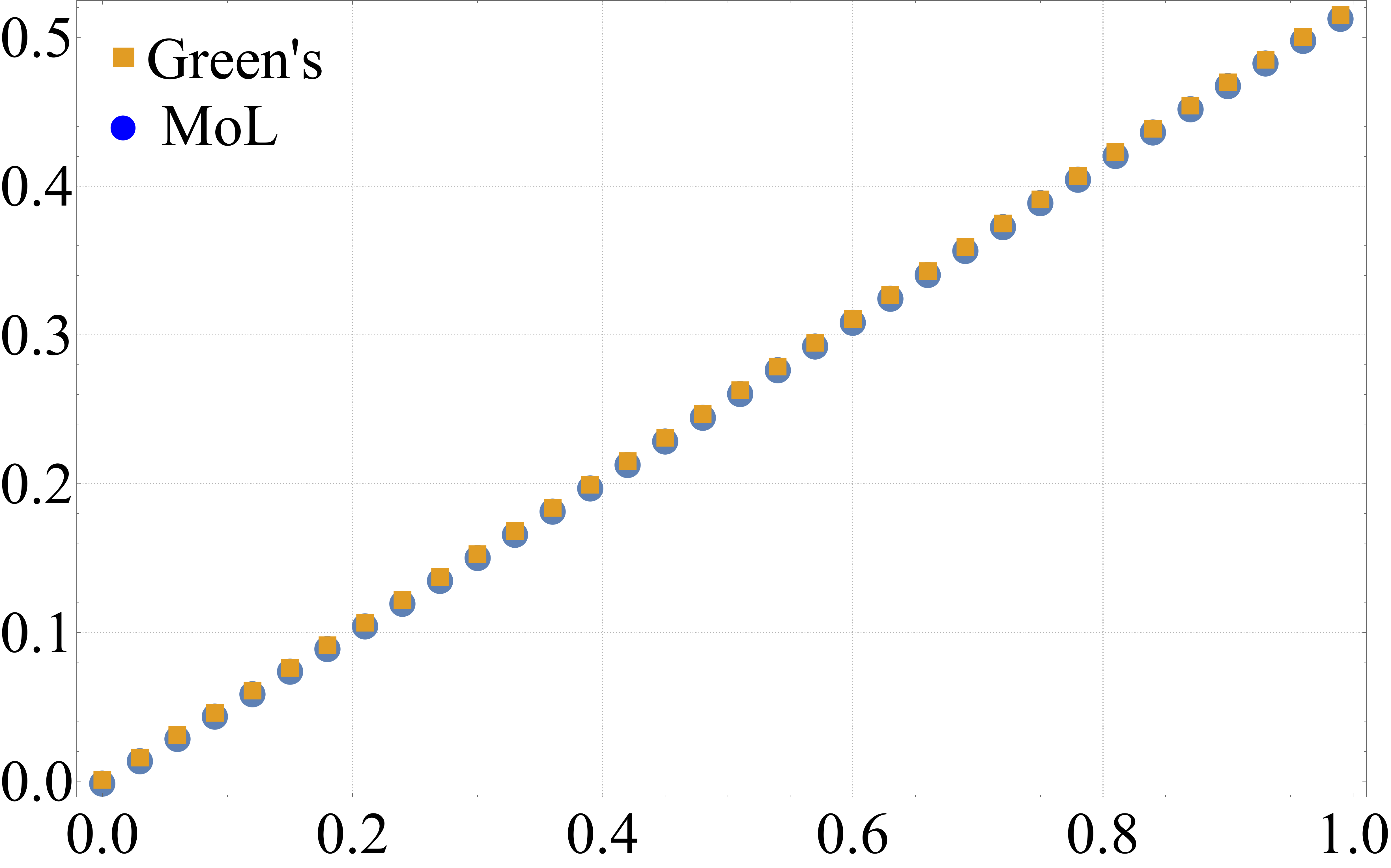} ~ \includegraphics[width=3.3in]{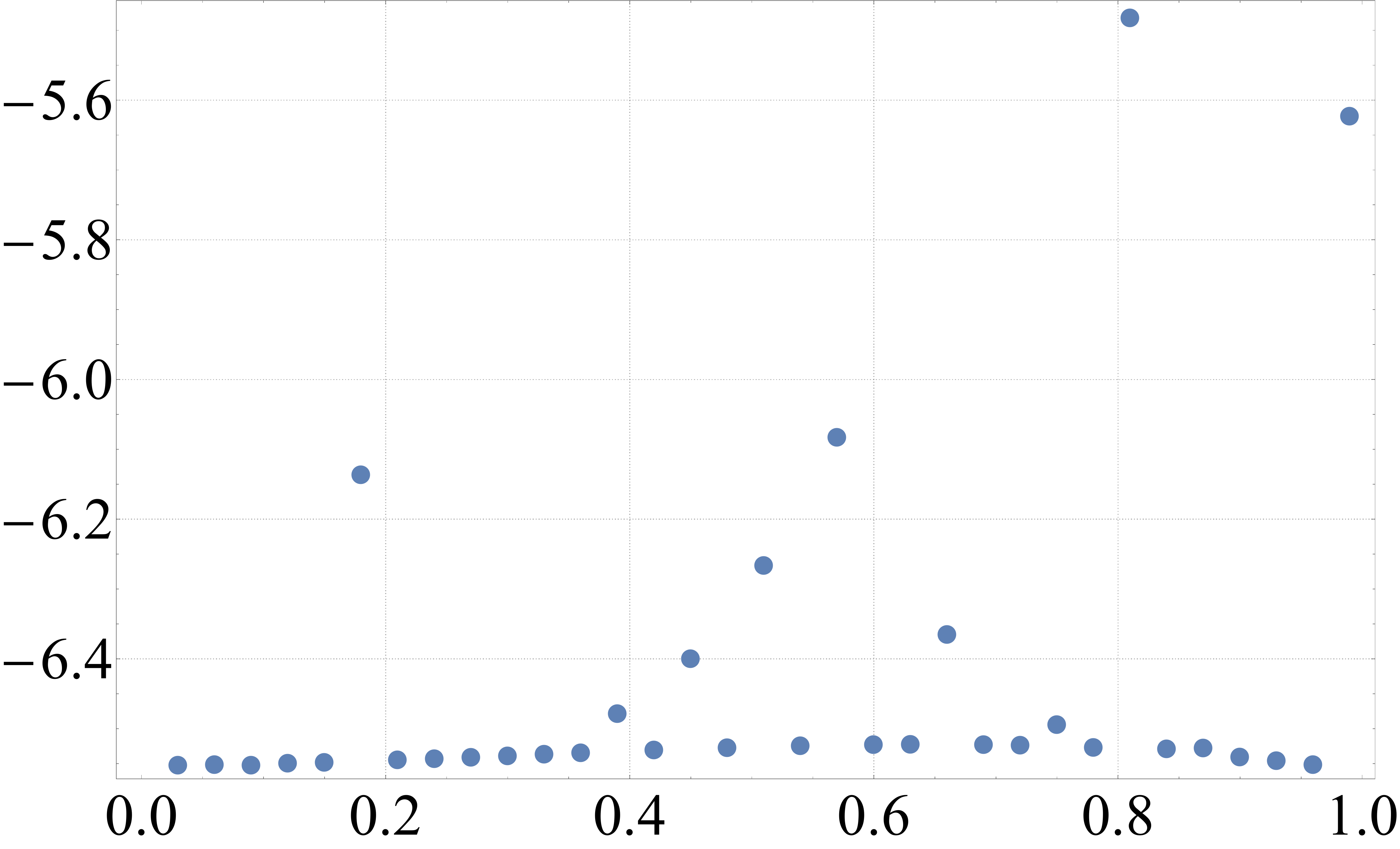}}
\caption{Discrete plot of $w_{\rm Green's}^1$ and $w_{\rm MoL}$ (left) and $\operatorname{Er}_1\left(\zeta; 1\right)$ (right) for $f\left(\zeta\right) = \delta\left(\zeta\right)$: KdV}
\label{fig1}
\end{figure}

We observe the error evolution of approximation by $w_{\rm Green's}^N$ of the Kortweg-de Vries equation above for different values of $N$. First, consider the case when $f\left(\zeta\right) = \delta\left(\zeta\right)$. This case allows to quantify how precisely does (\ref{GreenKdV}) approximate the Green's function obtained directly from (\ref{Greeneq}) by means of the method of lines. In Fig. \ref{fig1} we plot the two solutions and the error of approximation when $N = 1$.

\begin{figure}[H]
\centerline{\includegraphics[width = 3.2in]{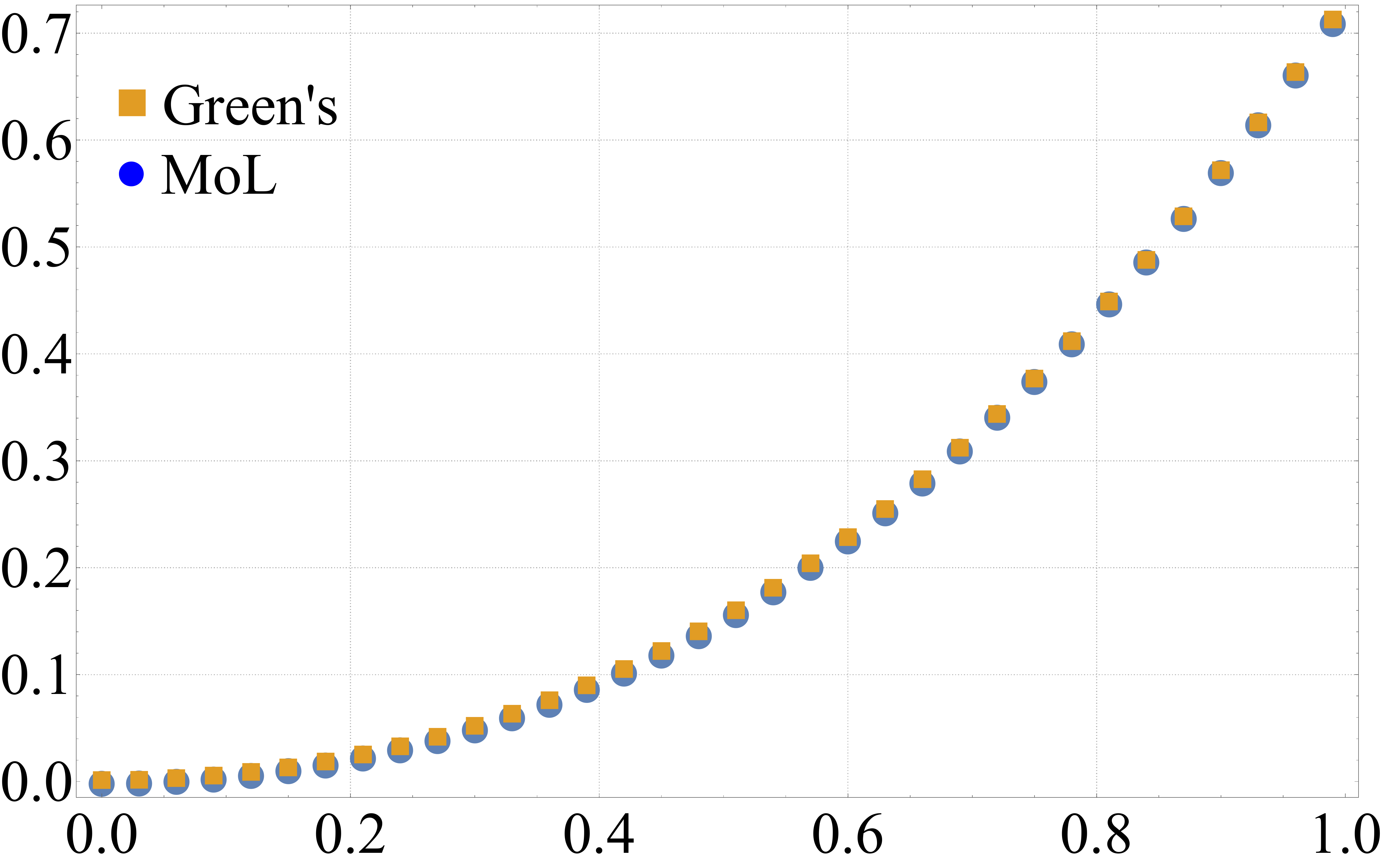} ~ \includegraphics[width=3.3in]{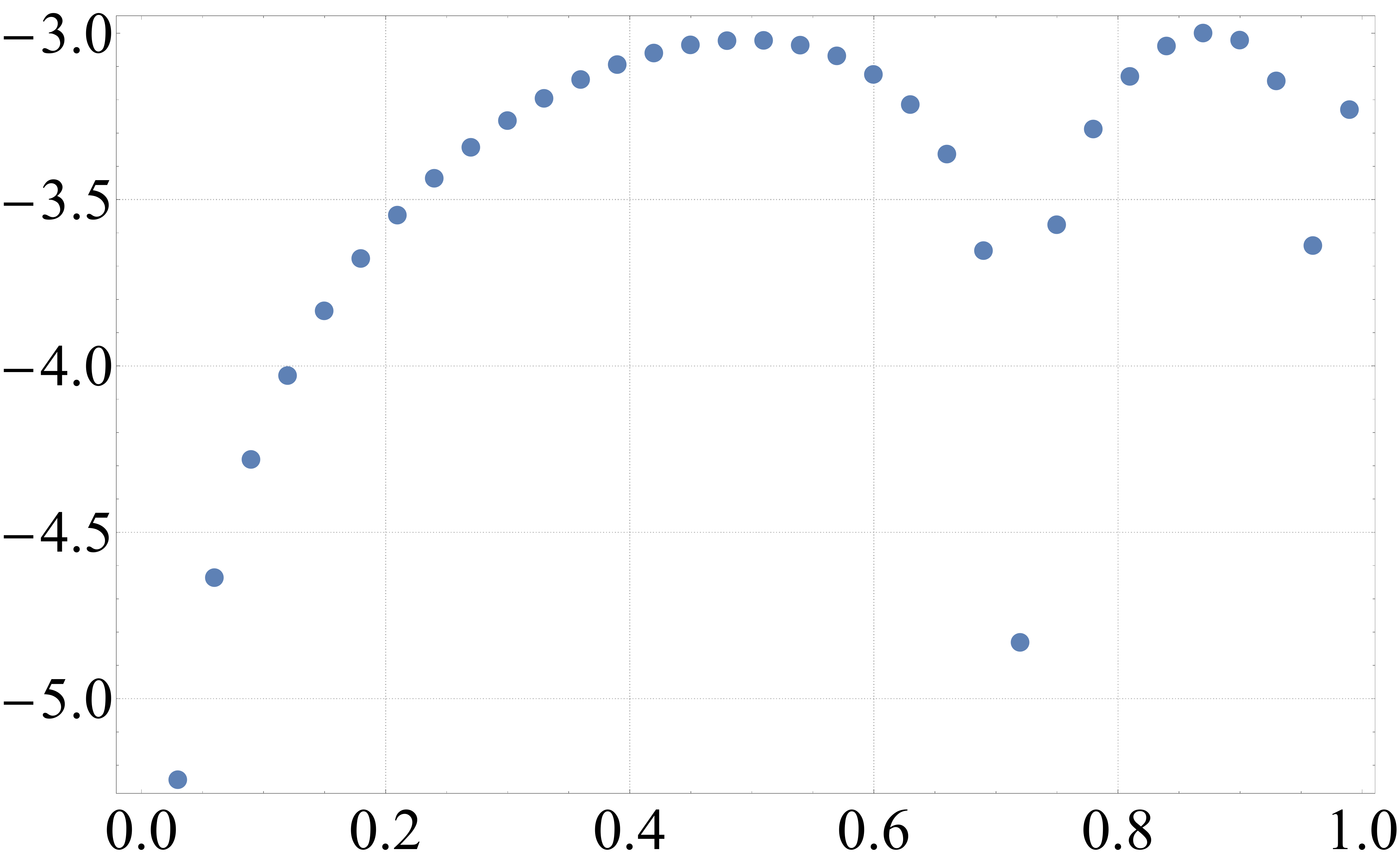}}
\caption{Discrete plot of $w_{\rm Green's}^1$ and $w_{\rm MoL}$ (left) and $\operatorname{Er}_1\left(\zeta; 1\right)$ (right) for $f\left(\zeta\right) = \exp\left(\zeta\right)$: KdV}
\label{fig2}
\end{figure}

\begin{figure}[H]
\centerline{\includegraphics[width = 3.3in]{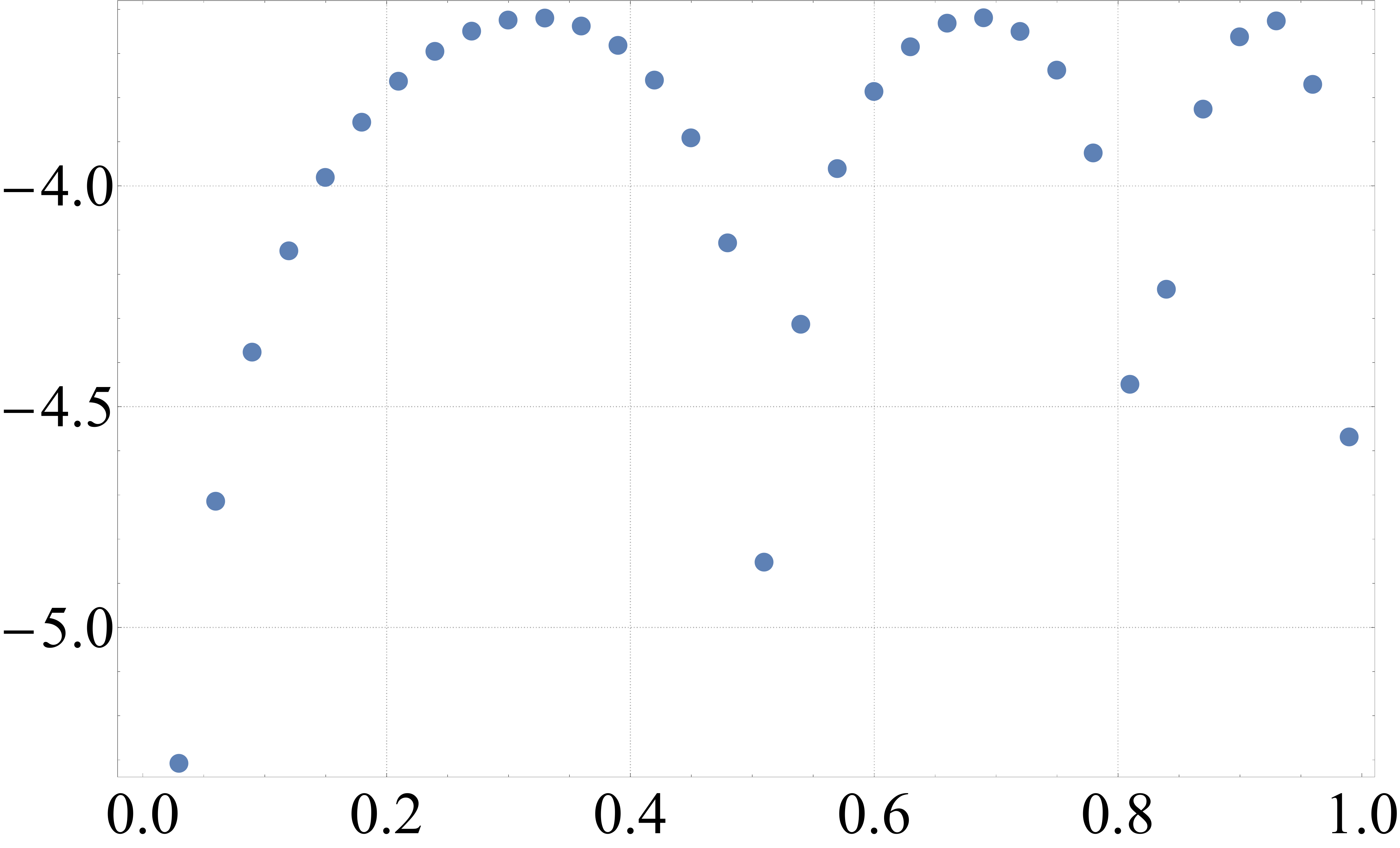} ~ \includegraphics[width=3.3in]{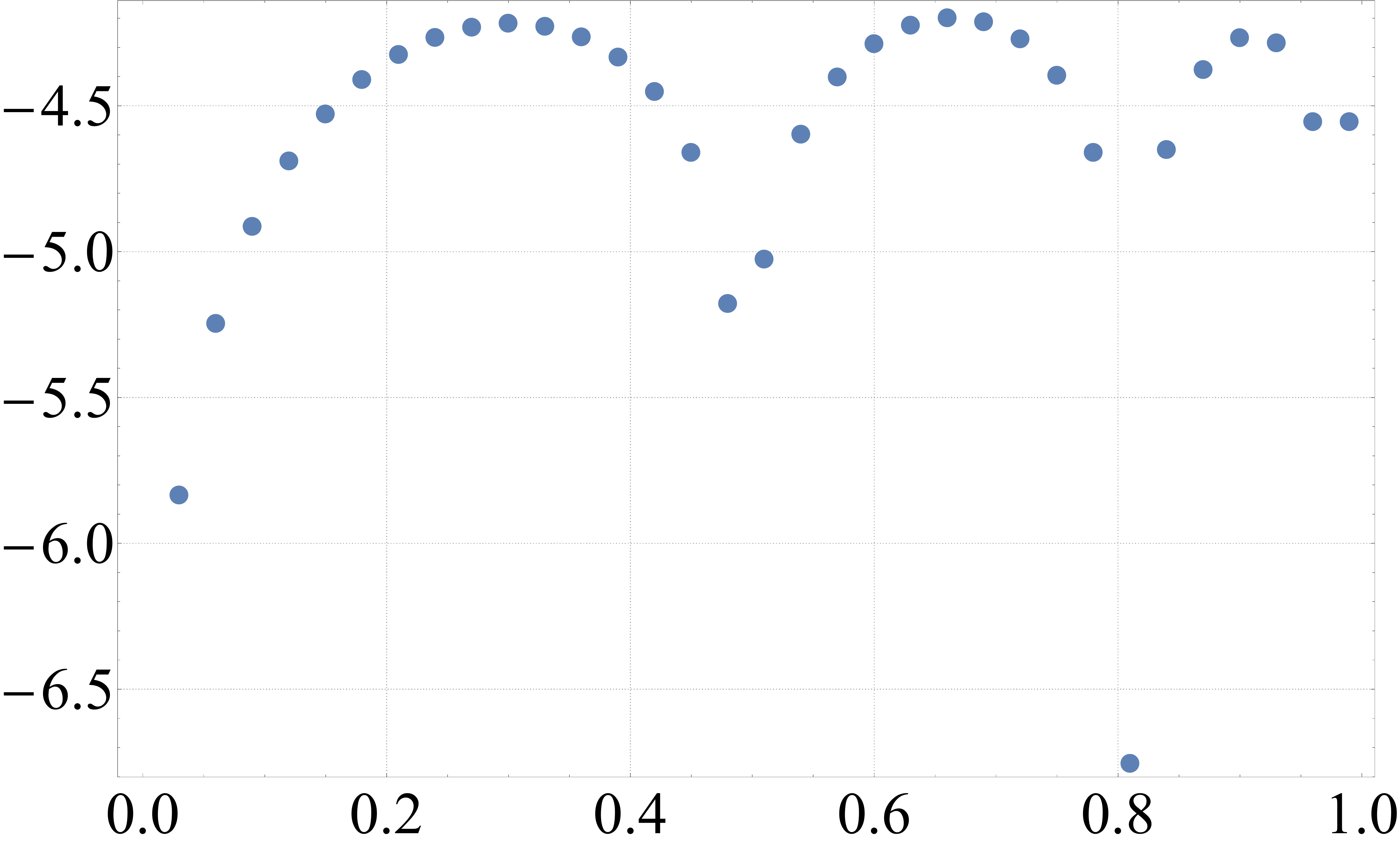}}
\caption{Discrete plot of $\operatorname{Er}_1\left(\zeta; 2\right)$ (left) and $\operatorname{Er}_1\left(\zeta; 4\right)$ (right) for $f\left(\zeta\right) = \exp \zeta$: KdV}
\label{fig3}
\end{figure}

Further, we study the influence of higher order terms on the approximation error when $f\left(\zeta\right) = \exp \zeta$. Fig. \ref{fig2} shows the logarithmic error of approximation by the first order term. Fig. \ref{fig3} shows how the error decreases when the number of terms increases. We gather the minimal and maximal logarithmic error for this case in Tab. \ref{tab1}.

\begin{table}[H]
\centering
\begin{tabular}{ c c c }
 $N$ & $\max\operatorname{Er}_1$ & $\min\operatorname{Er}_1$ \\ \hline
 
 1 & -3 & -5.3 \\ \hline
 
 2 & -3.6 & -5.3 \\ \hline
 
 4 & -4.25 & -6.6 \\ \hline

\end{tabular}
\caption{Minimal and maximal values of logarithmic error for different values of $N$ for $f\left(\zeta\right) = \exp \zeta$: KdV}
\label{tab1}
\end{table}

\subsection{Quadratic non-linearity}

Consider now the fourth order ODE (\ref{fourthorderODE}). As noted above, its Green's function has the form
\[
G\left(\zeta\right) = \theta\left(\zeta\right) \cdot w_0\left(\zeta\right),
\]
where $w_0$ satisfies (\ref{fourthorderODE}) and the following Cauchy conditions:
\[
w\left(0\right) = \frac{d w}{d \zeta}\bigg|_{\zeta = 0} = \frac{d^2 w}{d \zeta^2}\bigg|_{\zeta = 0} = 0, ~~ \frac{d^3 w}{d \zeta^3}\bigg|_{\zeta = 0} = s.
\]

Representing $w_0$ as a power series:
\[
w_0\left(\zeta\right) = \sum_{n = 0}^\infty \alpha_n \zeta^n,
\]
and taking into account that
\[
\frac{d^n w}{d \zeta^n}\bigg|_{\zeta = 0} = n! \alpha_n,
\]
from (\ref{fourthorderODE}) and the Cauchy conditions for the unknown coefficients $\alpha_n$ we derive
\[
\alpha_0 = \alpha_1 = \alpha_2 = 0, ~~ \alpha_3 = \frac{s}{6},
\]
\[
\left(n + 1\right) \left(n + 2\right) \left(n + 3\right) \left(n + 4\right) \alpha_{n + 4} - v^2 \beta_n \alpha_{n + 2} + \sum_{k = 0}^n \beta_k \alpha_{k + 2} \alpha_{n - k} = 0,
\]
where $\beta_n = \left(n + 1\right) \left(n + 2\right)$.

\begin{figure}[H]
\centerline{\includegraphics[width = 3.2in]{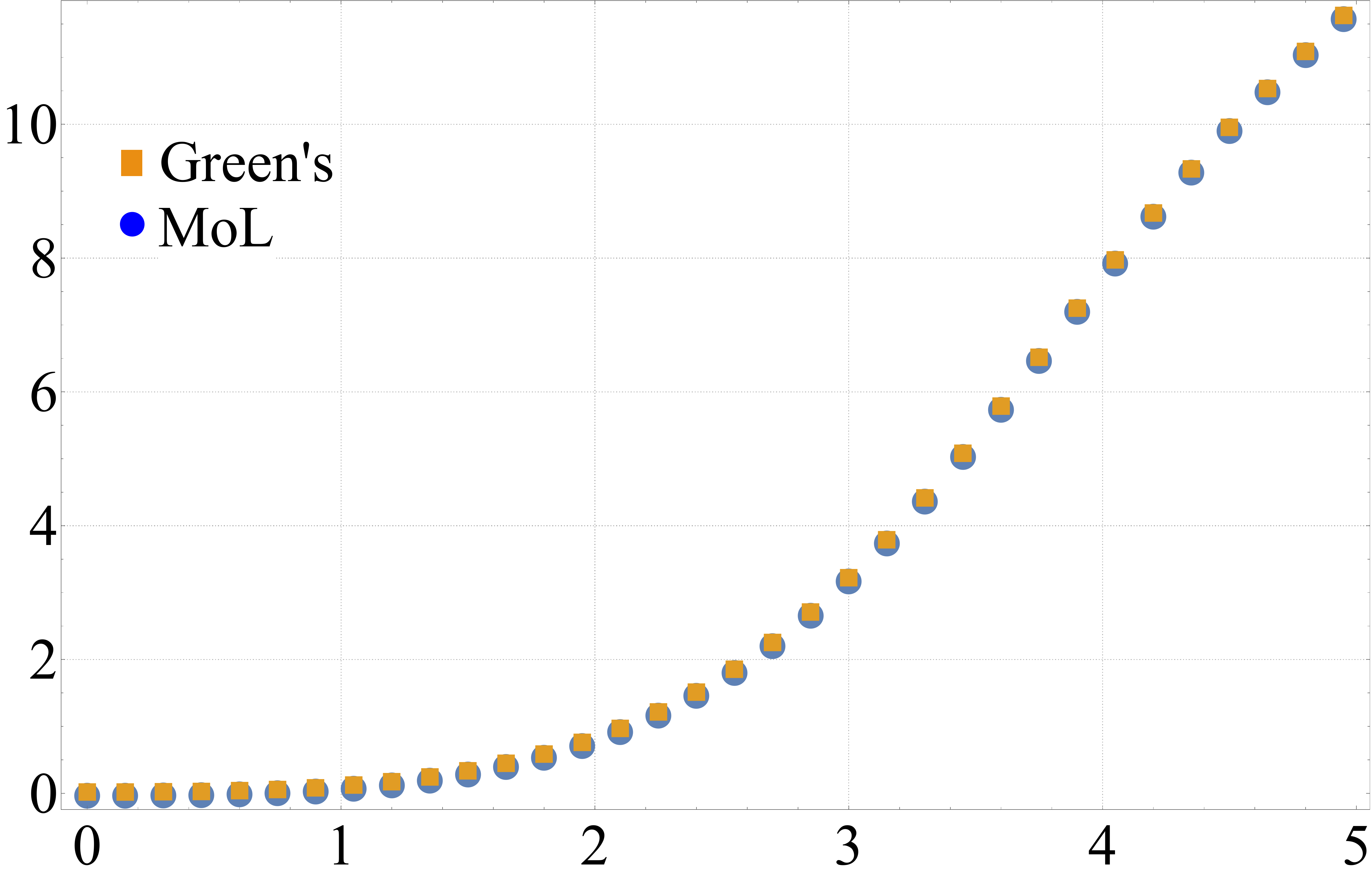} ~ \includegraphics[width=3.3in]{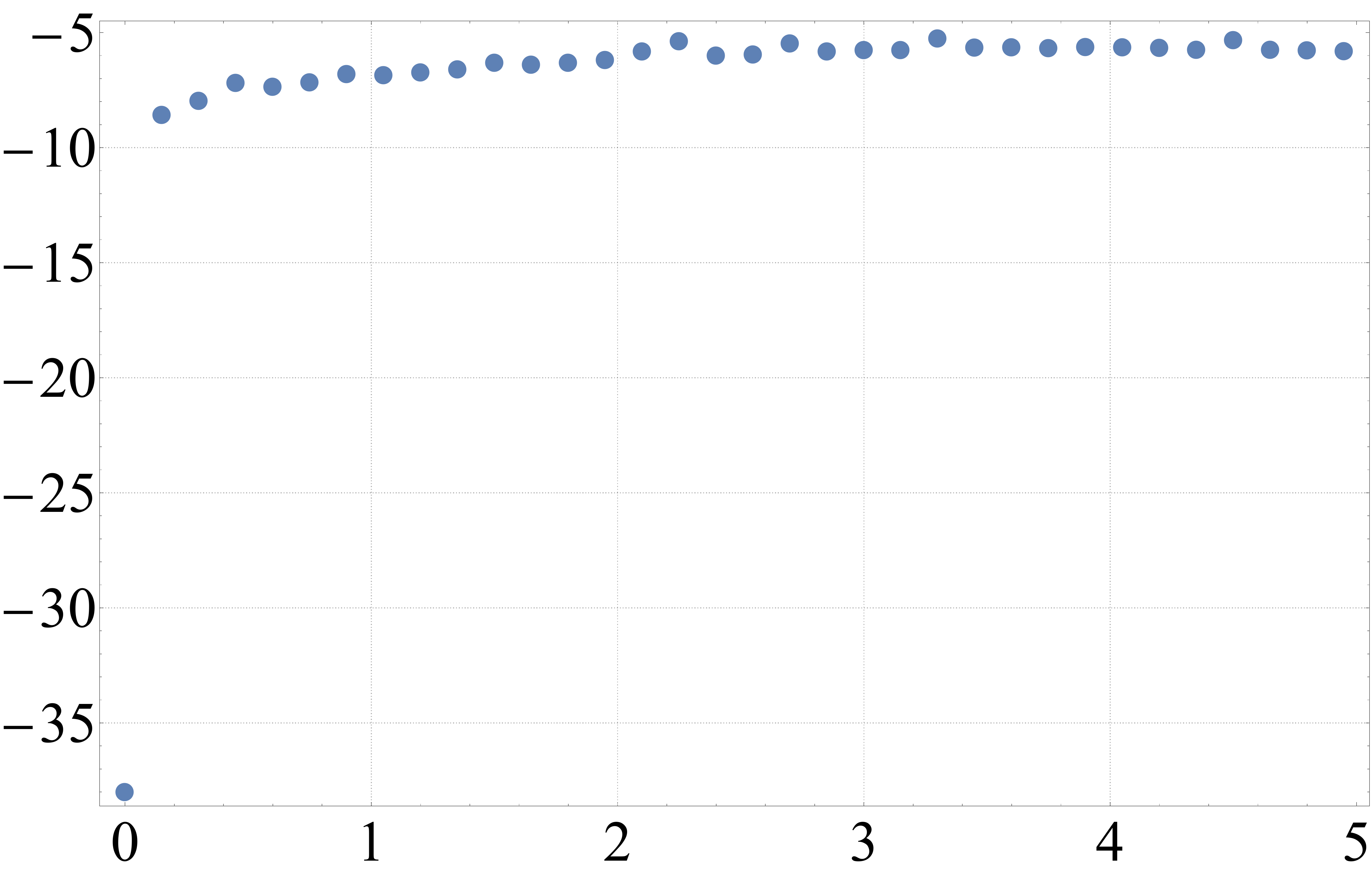}}
\caption{Discrete plot of $w_{\rm Green's}^1$ and $w_{\rm MoL}$ (left) and $\operatorname{Er}_1\left(\zeta; 1\right)$ (right) for $f\left(\zeta\right) = \delta\left(\zeta\right)$: (\ref{fourthorderODE})}
\label{fig4}
\end{figure}

We again consider the case when $f\left(\zeta\right) = \delta\left(\zeta\right)$, in order to quantify how accurate is the numerical calculation of the Green's function. Fig. \ref{fig4} allows to see that even the first order approximation of the Green's function solution in this case gives a low error approximation for nonlinear equations. As a matter of fact, in this case $\max\operatorname{Er}_1\left(\zeta; 1\right) \approx -5$.

\subsection{Boussinesq equation}

\begin{figure}[H]
\centerline{\includegraphics[width = 3.2in]{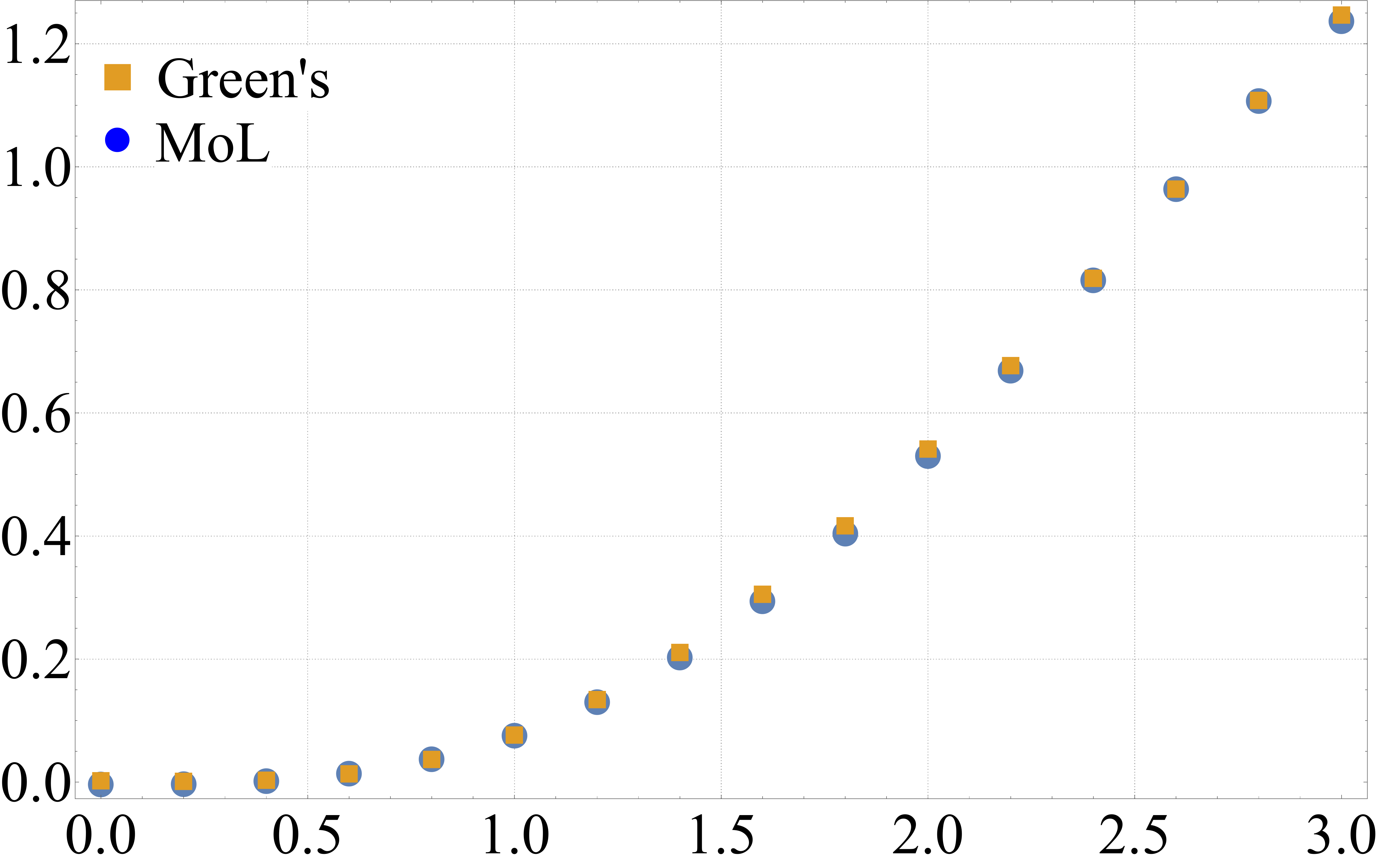} ~ \includegraphics[width=3.25in]{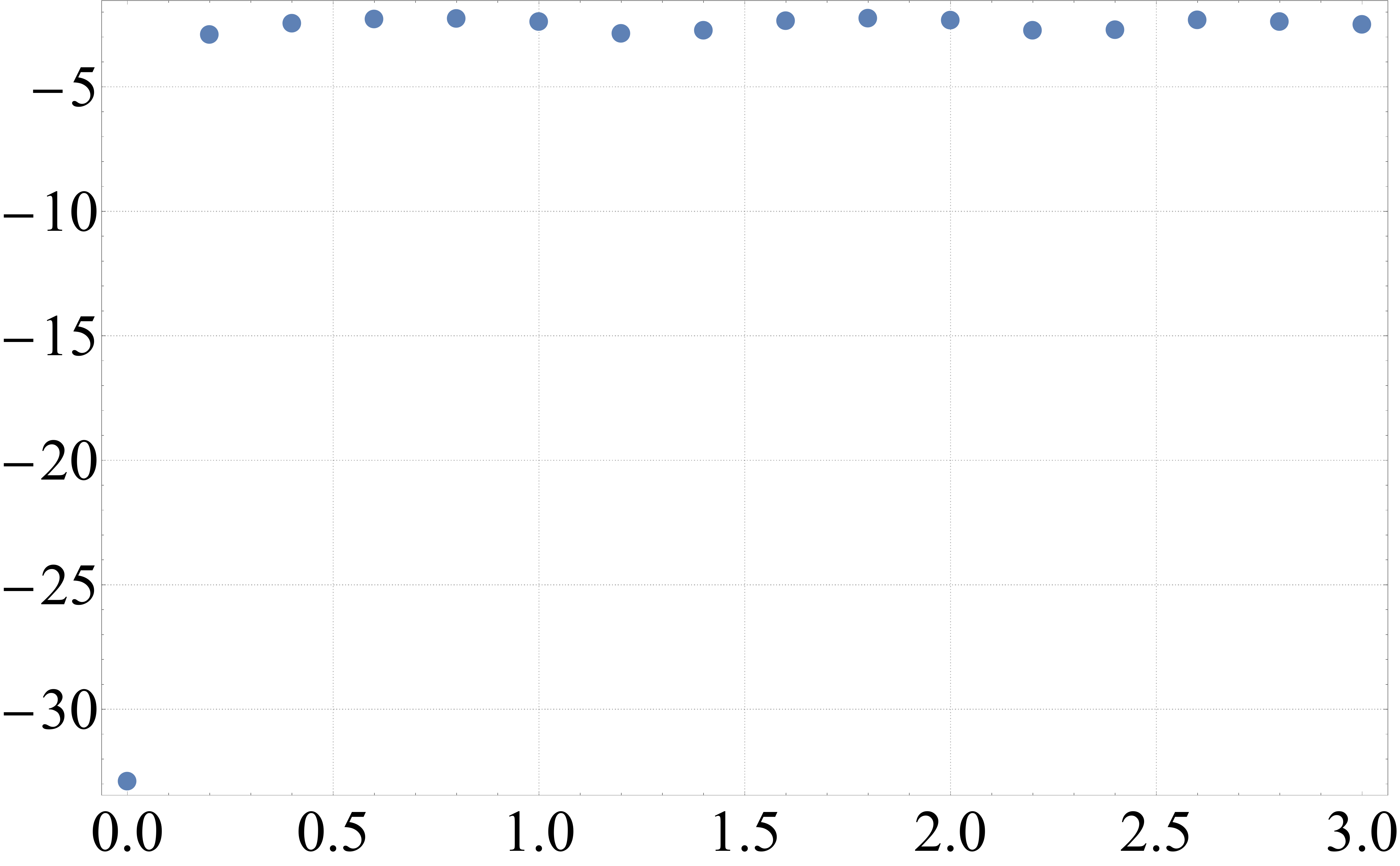}}
\caption{Discrete plot of $w_{\rm Green's}^3$ and $w_{\rm MoL}$ (left) and $\operatorname{Er}_1\left(\zeta; 3\right)$ (right) for $f\left(\zeta\right) = \delta\left(\zeta\right)$: Boussineq equation}
\label{fig5}
\end{figure}

In this subsection, we study the order one Green's function solution of the Boussinesq equation. First, consider the case when $f\left(\zeta\right) = \delta\left(\zeta\right)$. Fig. \ref{fig5} shows the approximation error for the solution of Boussinesq equation and $w_{\rm Green's}^3$. We see that even $N = 3$ ensures $\operatorname{Er}_1\left(\zeta; 3\right) \approx -5$.

\begin{figure}[H]
\centerline{\includegraphics[width = 3.2in]{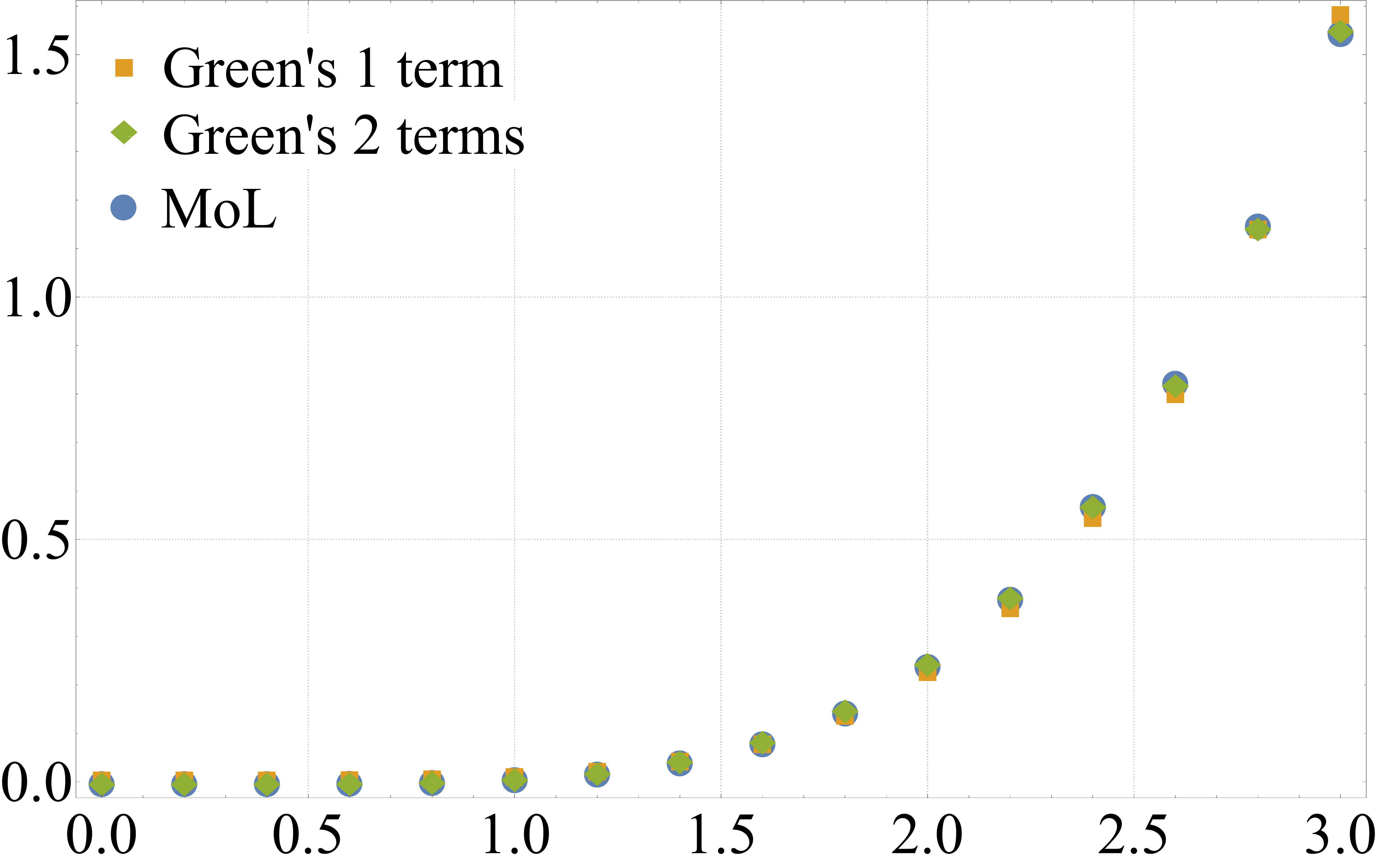}~\includegraphics[width = 3.2in]{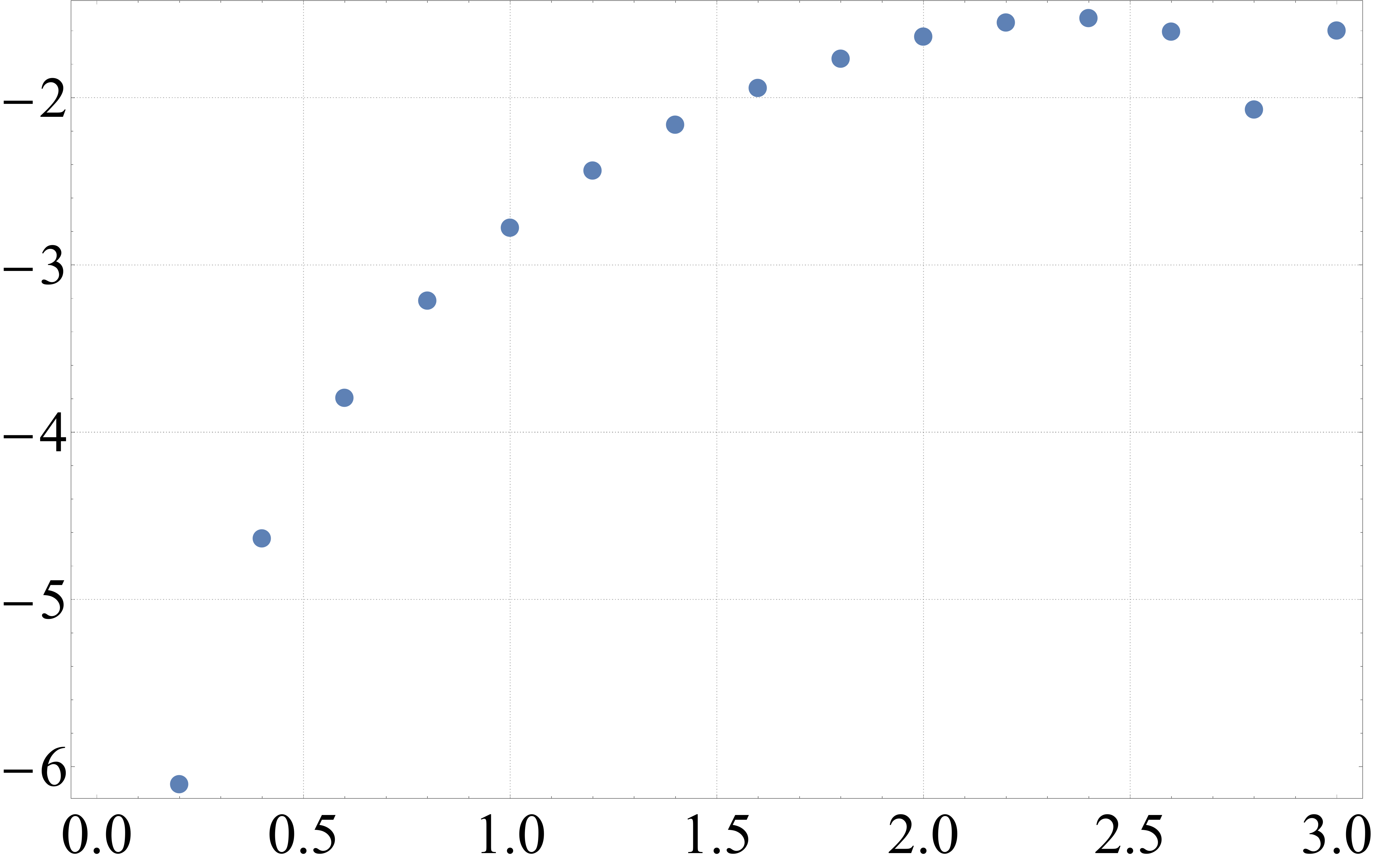}}
\caption{Discrete plot of $w_{\rm Green's}^1$, $w_{\rm Green's}^2$ and $w_{\rm MoL}$ (left) and $\operatorname{Er}_2\left(\zeta; 1, 2\right)$ (right) for $f\left(\zeta\right) = \zeta$: Boussinesq equation}
\label{fig6}
\end{figure}

Fig. \ref{fig6} shows the first ($N = 1$) and second ($N = 2$) order approximations of the Boussineq equation in the case when $f\left(\zeta\right) = \zeta$. We observe that the second order term of the short time expansion (\ref{shorttime}) brings a significant correction to the approximation error.

\begin{figure}[H]
\centerline{\includegraphics[width = 3.2in]{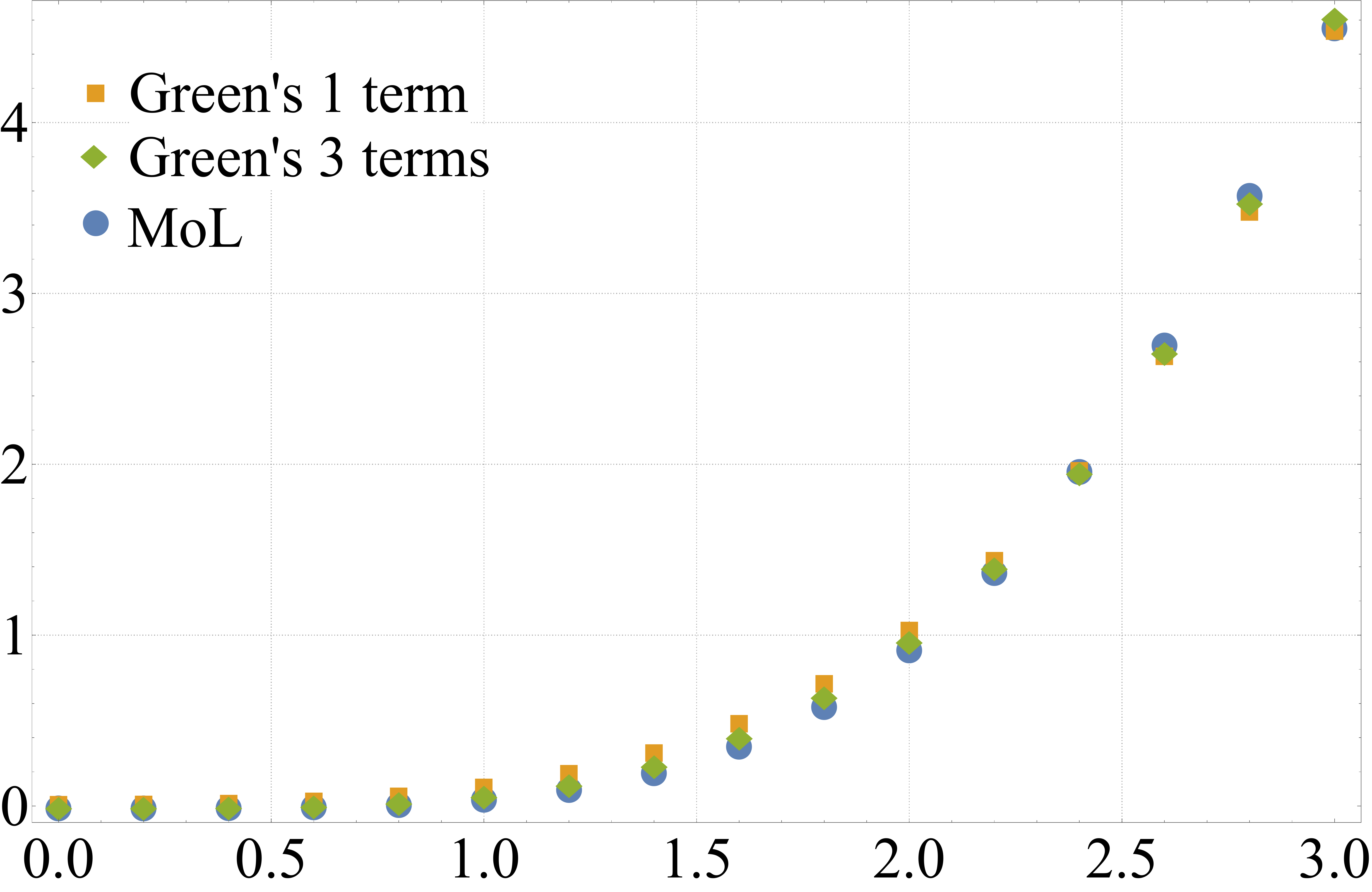}~\includegraphics[width = 3.35in]{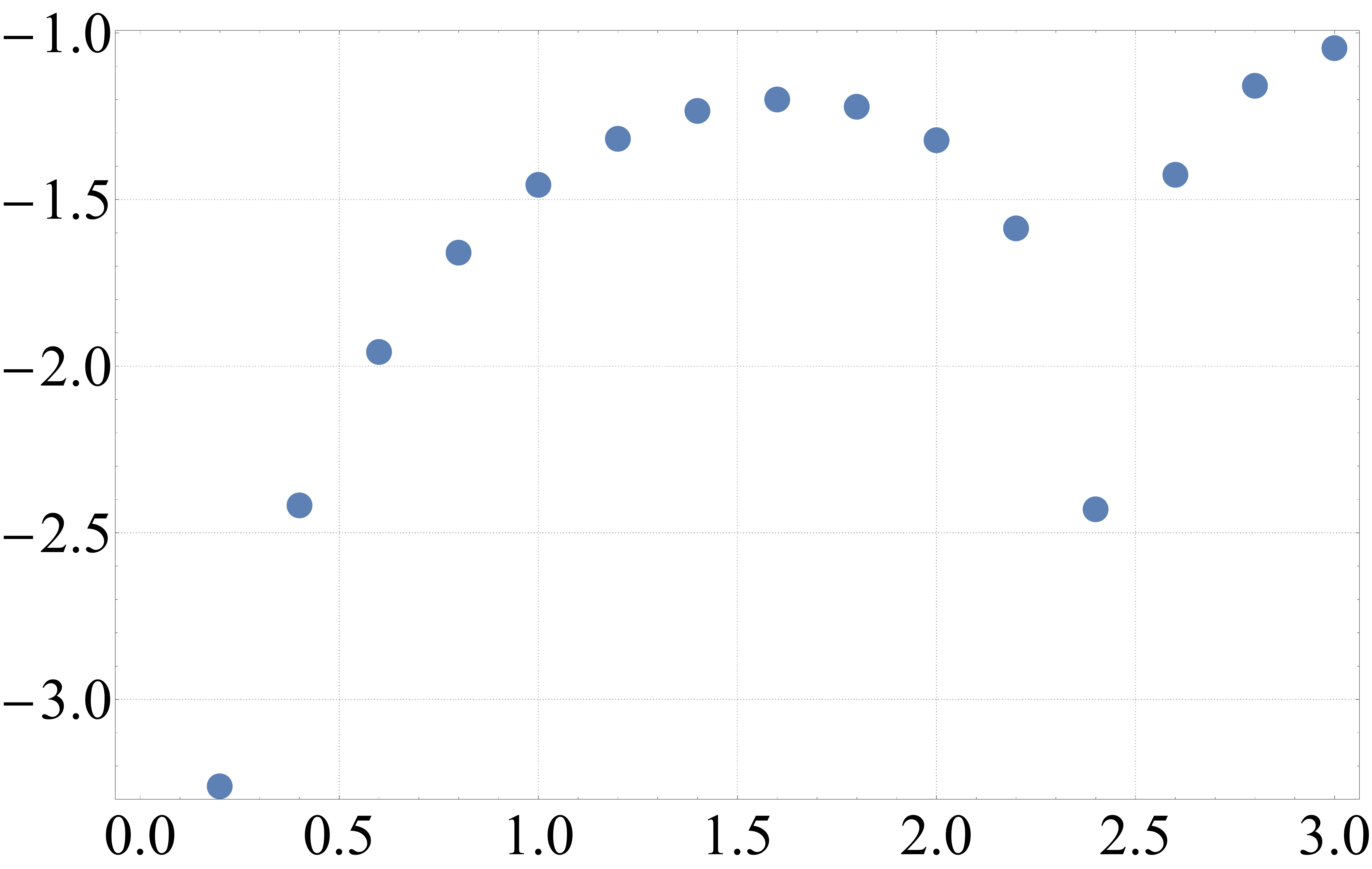}}
\caption{Discrete plot of $w_{\rm Green's}^1$, $w_{\rm Green's}^3$ and $w_{\rm MoL}$ (left) and $\operatorname{Er}_2\left(\zeta; 1, 3\right)$ (right) for $f\left(\zeta\right) = \exp \zeta$: Boussinesq equation}
\label{fig7}
\end{figure}

\begin{figure}[H]
\centerline{\includegraphics[width = 3.2in]{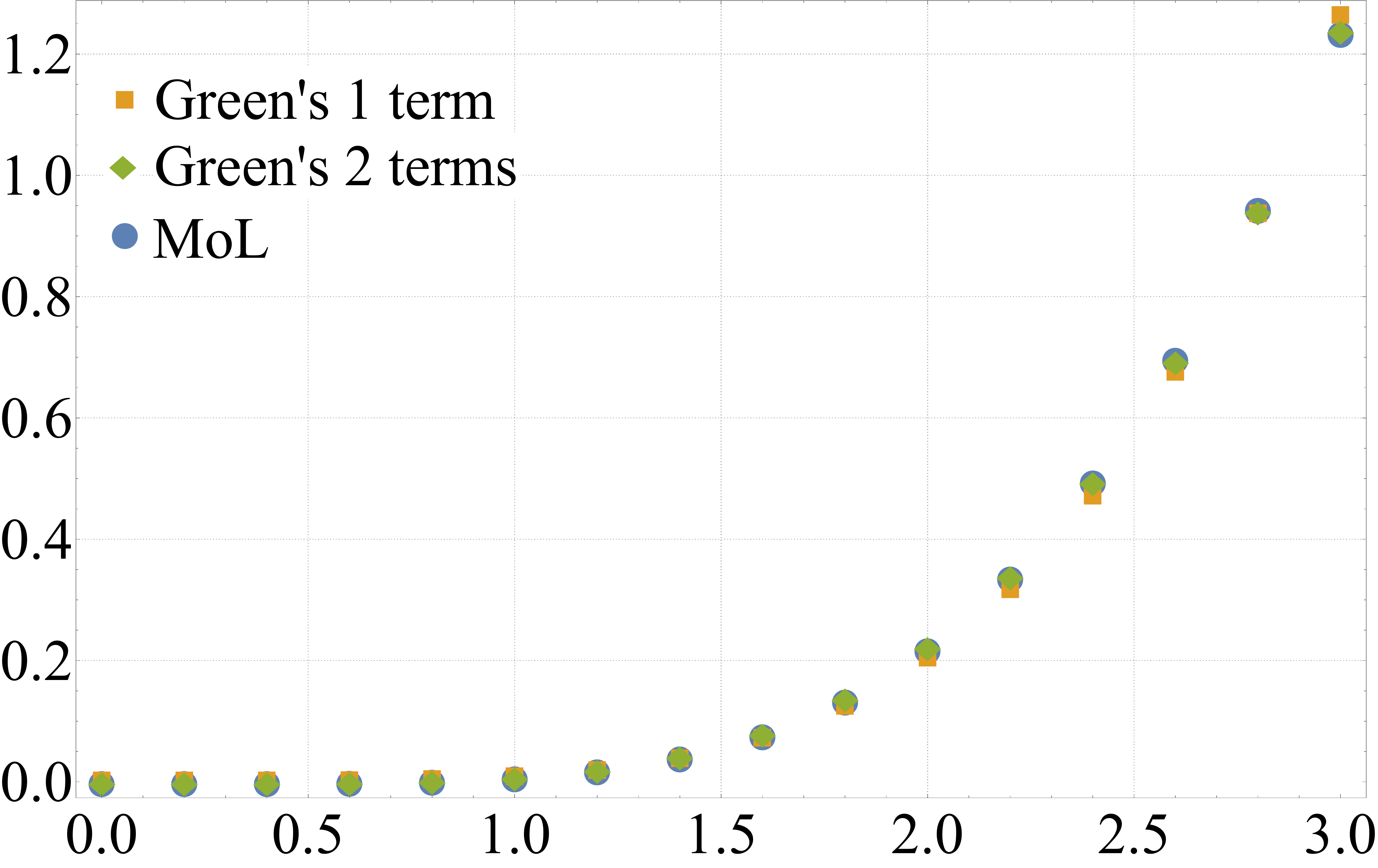}~\includegraphics[width = 3.2in]{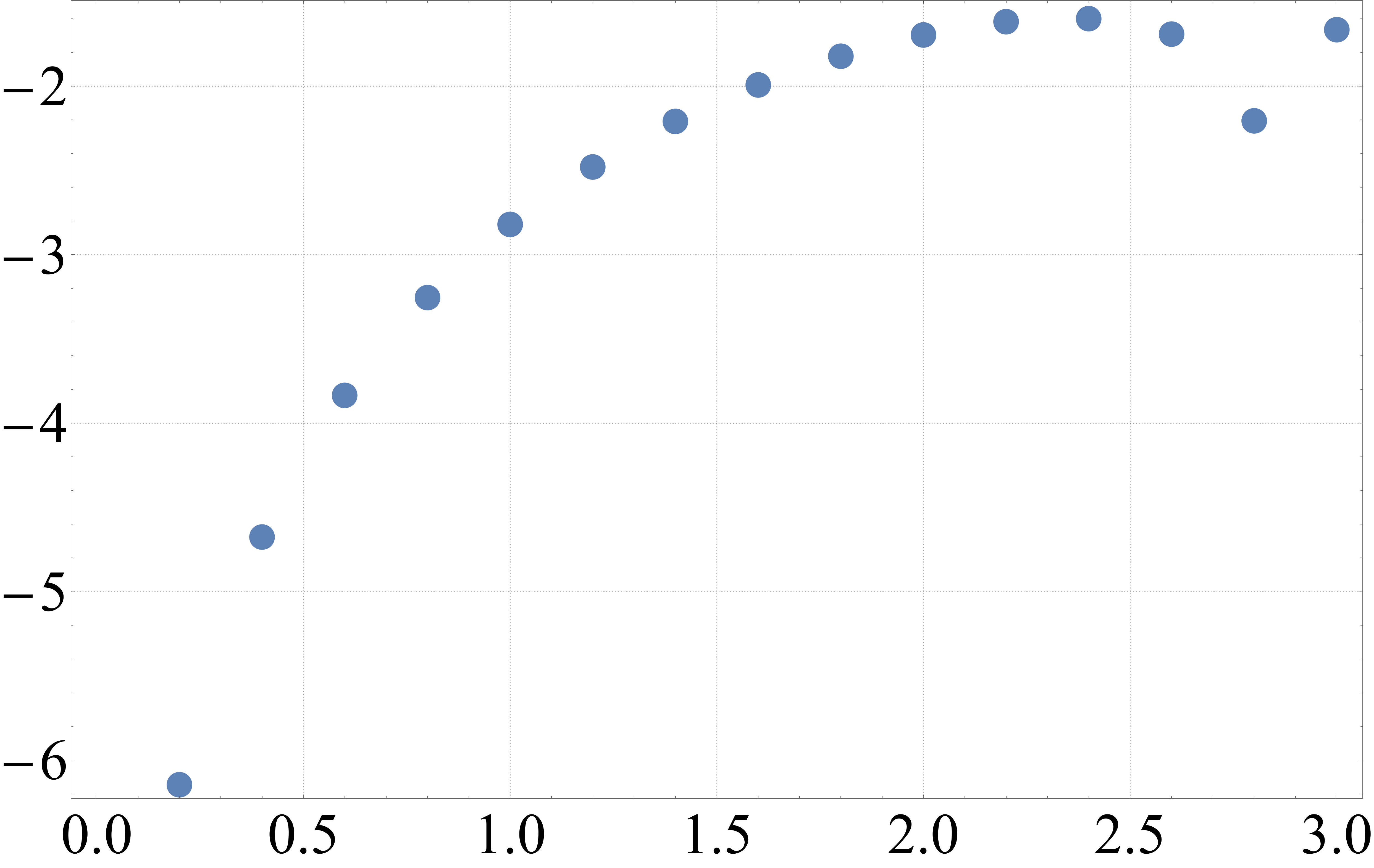}}
\caption{Discrete plot of $w_{\rm Green's}^1$, $w_{\rm Green's}^2$ and $w_{\rm MoL}$ (left) and $\operatorname{Er}_2\left(\zeta; 1, 2\right)$ (right) for $f\left(\zeta\right) = \sin \zeta$: Boussinesq equation}
\label{fig8}
\end{figure}

\begin{figure}[H]
\centerline{\includegraphics[width = 3.2in]{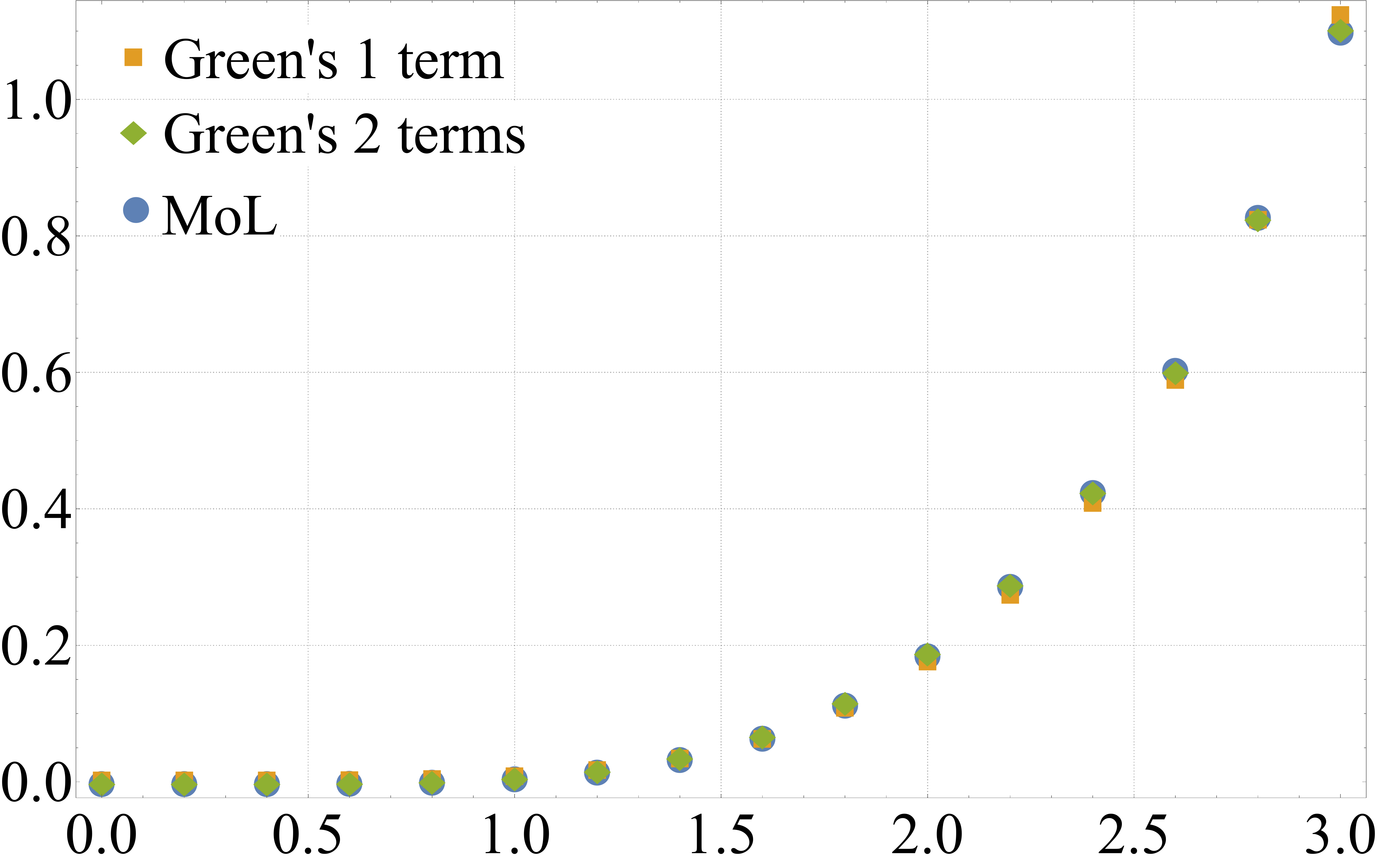}~\includegraphics[width = 3.3in]{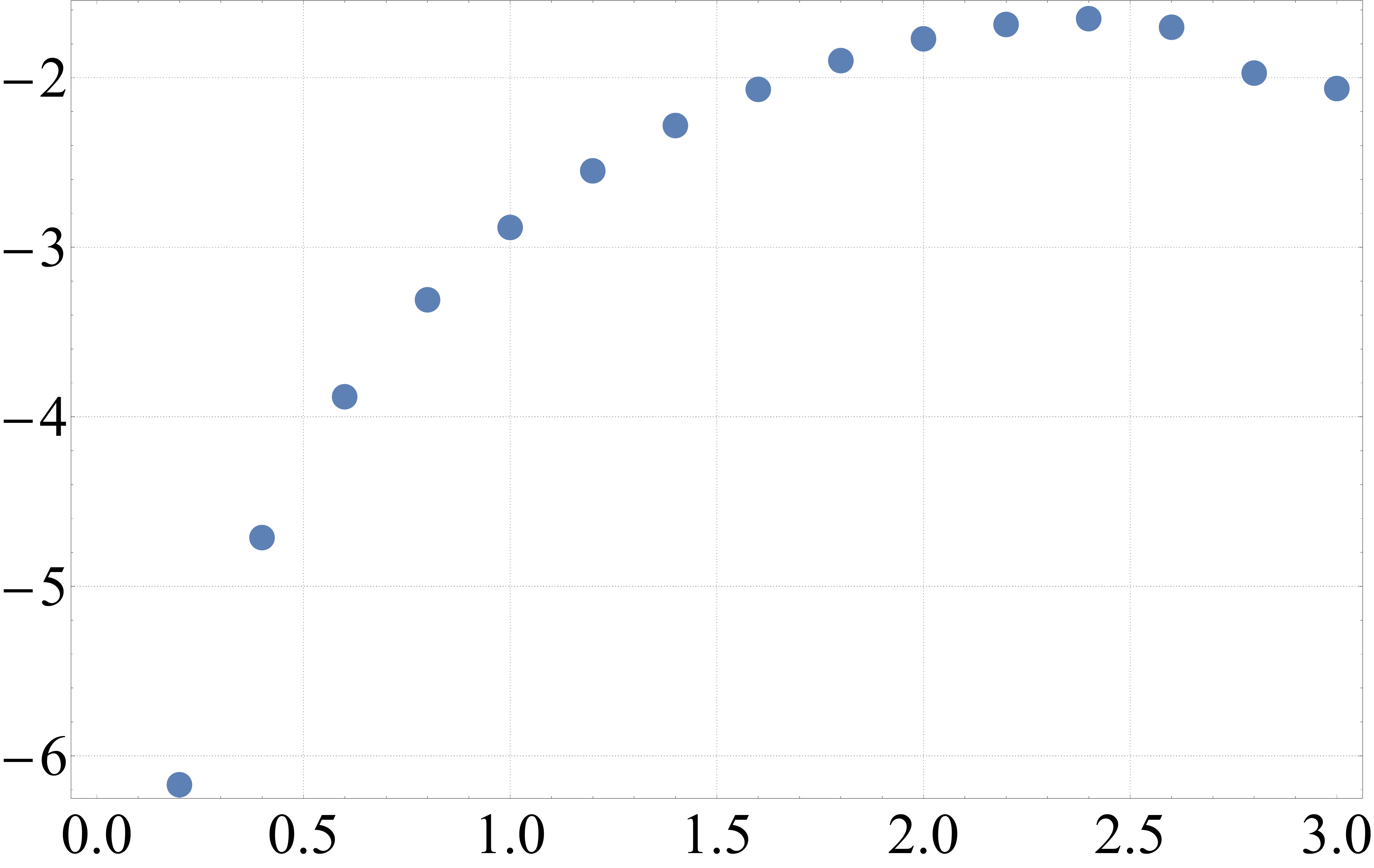}}
\caption{Discrete plot of $w_{\rm Green's}^1$, $w_{\rm Green's}^2$ and $w_{\rm MoL}$ (left) and $\operatorname{Er}_2\left(\zeta; 1, 2\right)$ (right) for $f\left(\zeta\right) = \ln\left(1 + \zeta\right)$: Boussinesq equation}
\label{fig9}
\end{figure}

Numerical analysis reveals a sort of robustness of the the Green's function solution with respect to the source term $f$. Indeed, as Figs. \ref{fig7}-\ref{fig9} show, the short time expansion (\ref{shorttime}) for the Boussinesq equation provides low error semi-analytical approximation.

As shows the evaluation of the error function $\operatorname{Er}_2$, the main approximation is carried out by $w_{\rm Green's}^1$, while higher order terms contribute to the error reduction.

\section*{Conclusion}

In this paper we show that, if the nonlinear term of an $n$th order differential equation which is linear in the highest order derivative, is a homogeneous function in the above sense, then its nonlinear Green's function is represented as a product of the Heaviside $\theta$ function and the homogeneous solution satisfying non-homogeneous boundary conditions. We support the theoretical derivations by specific examples including KdV and Boussinesq equations. Numerical error analysis shows a good correspondence of the method we discuss here with the well-known numerical method of lines. Furthermore, the contribution of the higher order terms of the short time expansion is studied for different source terms. The main advantage of this method is that the derived solution depends on the source term and initial conditions \emph{explicitly} making rigorous analysis of nonlinear systems much simpler.

\end{document}